\PassOptionsToPackage{fleqn}{amsmath}
\documentclass[twocolumn,preprintnumbers,amsmath,amssymb,floatfix]{revtex4}
\usepackage{graphicx}
\usepackage{dcolumn}
\usepackage{bm}
\usepackage{amssymb}
\usepackage{amsmath}

\begin{document}
\title{Anisotropy and interaction effects of strongly strained SrIrO$_3$ thin films}

\author{L. Fruchter$^{1}$, O. Schneegans$^{2}$, Z.Z. Li$^{1}$}%
\affiliation{$^1$Laboratoire de Physique des Solides, C.N.R.S., Universit\'{e} Paris-Sud, 91405 Orsay, France}
\affiliation{$^2$Laboratoire G\'{e}nie Electrique et Electronique de Paris, CNRS UMR 8507, UPMC and Paris-Saclay Universities, Centralesup\'{e}lec, 91192 Gif Sur Yvette, France}
\date{Received: date / Revised version: date}

\begin{abstract}{Magneto-transport properties of SrIrO$_3$ thin films epitaxially grown on SrTiO$_3$, using reactive RF sputtering, are investigated. A large anisotropy between the in-plane and the out-of-plane resistivities is found, as well as a signature of the substrate cubic to tetragonal transition. Both observations result from the structural distortion associated to the epitaxial strain. The low-temperature and field dependences of the Hall number are interpreted as due to the contribution of Coulomb interactions to weak localization, evidencing the strong correlations in this material. The introduction of a contribution from magnetic scatters, in the analysis of magnetoconductance in the weakly localized regime, is proposed as an alternative to an anomalously large temperature dependence of the Land\'{e} coefficient.}
\end{abstract}
\maketitle

\section*{\label{intro}Introduction}

The Ruddlesden-Popper series, R$_{n+1}$Ir$_n$O$_{3n+1}$ where R= Sr, Ba and n = 1,2,$\infty$, has emerged as a new playground for the study of electron correlation effects. In these compounds, while extended 5$\textit{d}$ orbitals tend to reduce the electron-electron interaction, as compared to the 3$\textit{d}$ transition metal compounds as cuprates, the strong spin orbit coupling (SOC) associated to the heavy Ir and the on-site Coulomb interaction compete with electronic bandwidth to restore such correlations\cite{Kim08}. As a result, electronic properties are determined by the balance between charge, spin, and correlation, and this delicate balance can be tailored by tuning of the structure of these compounds. In the series, SrIrO$_3$ (n = $\infty$, SIO) is metallic and it was shown that, opposite to common wisdom, a magnetic metal is obtained with a larger SOC at a larger electron-electron interaction\cite{Zeb12}. This unusual behavior is thought to result from the small density of states near the Fermi level, requiring a large interaction strength to open a gap. Contrasting with Sr$_2$IrO$_4$, SIO presents IrO$_2$ octahedra rotations both along (001) and (110), which explains the presence of narrower bands, in place of larger ones for increased coordination\cite{Nie15}. SIO has early attracted much attention of the thin film community\cite{Liu05,Kim06,Zhang13,Xiang13,Liu13,Gruenewald14,Zhang15,Biswas15}, as single crystals of the orthorhombic phase\cite{Longo71,Zhao08} may only be grown using epitaxy. It has appeared quickly that the sensitivity of the electronic properties to structure is here especially important\cite{Xiang13,Liu13,Gruenewald14,Zhang15}. Systematic thin film deposition on various substrates showed that in-plane lattice compression (alt. extension) is associated to an increase (alt. a decrease) of the resistivity, and to low temperature localization\cite{Biswas14,Gruenewald14}. On GdScO$_3$, which induces a small lattice extension, a recent study finds that strain may significantly alter the band structure of thin films, in particular opening a gap in place of the predicted symmetry-protected line nodes for SIO\cite{Liu16}. A similar opening was found, from band calculations and photoemission spectroscopy, for films on SrTiO$_3$, which induces a large lattice compression\cite{LiuLi15}. Despite such an attention, it remains unclear how correlations in this semimetal actually affect the transport properties. It was proposed that non-Fermi physics is evidenced by anomalous temperature power law behavior, and that electron-electron correlations contribute to the metal-insulator transition for the strained material\cite{Biswas14,Biswas16}. Also, the possible contribution of localized magnetic moments at the metal-insulator transition was stressed\cite{Biswas16}. We believe that such proposals however still need experimental strengthening, in particular from an appropriate interpretation of the magneto-transport properties. Here, we investigate these properties in detail, for SIO thin films grown on SrTiO$_3$. We put into evidence the strong contribution of the film strain to the anisotropic transport properties; we also show how magnetic scattering could account for the anomalous temperature dependence of the magnetoconductance, and revisit the temperature and field dependence of the Hall number, considering the electronic correlations.

\section{\label{structure}Samples}

\begin{table}
\caption{\label{tab:table1}SrIrO$_3$ thin films used in this study}
\begin{ruledtabular}
\begin{tabular}{cccc}
Sample&Substrate&Thickness (nm)&R.C. (deg.)\footnote{(220) rocking curve FWHM}\\
\hline
\textsl{A} & STO conv. & 8 & 0.06\\
\textsl{B} & STO conv.& 14 & 0.05\\
\textsl{C} & STO, miscut $\simeq$ 0.5 deg. & 34\\
\textsl{D} & STO vicinal 5 deg. & 37 & 0.06\\
\textsl{E} & STO vicinal 5 deg. & 37\\
\end{tabular}
\end{ruledtabular}
\end{table}

\begin{figure}
\resizebox{0.95\columnwidth}{!}{%
  \includegraphics{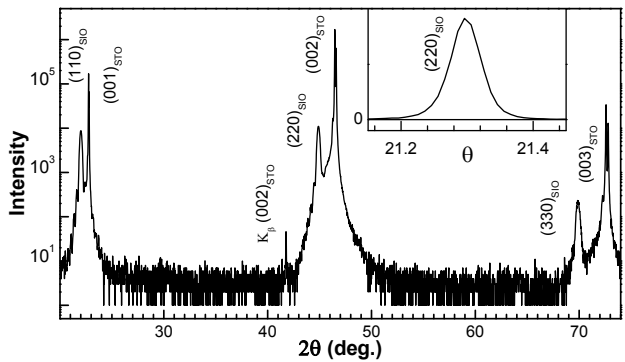}
  }
\caption{$\theta-2\theta$ diffraction scan for sample \textsl{C}. The inset is the (220) rocking curve.}\label{diffraction}
\end{figure}

\begin{figure}
\resizebox{0.95\columnwidth}{!}{%
  \includegraphics{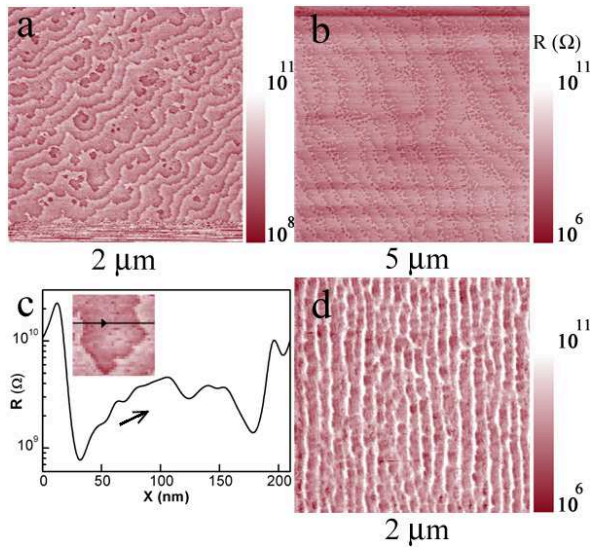}
  }
\caption{Conductive AFM scans. a) sample \textsl{A}; b) sample \textsl{B}; c) Resistance profile across two alternate steps in a), showing the resistance dips at the terrace steps; d) sample \textsl{D} - a close inspection shows there are several unit cells across each bright terrace step.}\label{cafm}
\end{figure}

Single phase epitaxial thin films of SIO on (001) SrTiO$_3$ substrates were synthesized using RF magnetron sputtering deposition, in oxygen/argon plasma, from an off-stoichiometry target. A detailed description of the synthesis is given elsewhere\cite{Li16}. The quality of the epitaxy was checked, using X-ray diffraction analysis. $\theta$-$2\theta$ were basically used to check for the absence of any parasitic phase (Fig.~\ref{diffraction}. Reflectivity modeling yielded a typical 0.4 nm roughness for the thin films, while AFM images revealed two-dimensional growth with steps of the same height (Fig.~\ref{cafm}). Rocking curves exhibited typical angular width 0.05-0.06 deg. for the (002) reflection, and the presence of Kissig fringes indicated \textit{c}-axis coherence. Two-dimensional reciprocal lattice scans in the vicinity of the (116) reflection revealed unrelaxed epitaxy of single-domain films. 

\begin{figure}
\resizebox{0.5\columnwidth}{!}{%
  \includegraphics{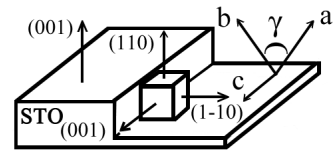}
  }
\caption{Sketch of the epitaxial growth, in the presence of a miscut step.}\label{epitaxy}
\end{figure}

The films were found oriented with the (110) direction perpendicular to the substrate, and the (001) direction was systematically aligned either along the (100) or (010) direction of the substrate. Further, systematic AFM investigations coupled to the X-ray diffraction analysis showed that the orientation of the substrate terraces (originating from the unavoidable miscut of the substrate) selects between these two orientations, in order to minimize the angle between the crystal (001) direction and the substrate steps. Such a result was anticipated in Ref.~\onlinecite{Zhang13}. The large lattice mismatch between the material (3.953 \AA) and the substrate (3.903 \AA) induces a large bi-axial compressive strain. In Ref.~\onlinecite{Liu16} (see also Ref.~\onlinecite{Vailionis08} for Sr(Ca)RuO$_3$), it was found that the substrate strain not only modifies the lattice parameters, but also the unit cell angles, lowering the symmetry to monoclinic for these two cases. For SIO on GdScO$_3$, the monoclinic distorsion is small ($\gamma \simeq$ 90.4 deg), but the oxygen octahedra are rotated by a few deg, towards the undistorted perovskite structure, with respect to the pristine crystal\cite{Liu16}. In the present case, with a lattice mismatch $\simeq$ 1.5 \%, the unit cell is expected to be strongly distorted, with a \textit{decrease }of the cell angle $\delta\gamma$ $\simeq$ 1.6 deg (Fig.~\ref{epitaxy}). Using (400), (040), (220) and (224) X-ray reflections, we obtained $\gamma$ = 88.1 deg., demonstrating that a strongly monoclinic cell is indeed a reasonable assumption in the present case. 
For magneto-transport studies, gold contacts where sputtered on the films, which were then patterned, using photolithography, in a standard Hall bar geometry (0.3 x 1 mm$^2$) or two perpendicular resistivity bars for anisotropy measurements (80 x 900 $\mu$m$^2$). The films were found totally insensitive to strong acid chemical etching (including hot \textit{aqua regia}), and mechanically very hard. So they were not affected by the photolithography process, as we could check from the comparison of resistivity on raw and patterned samples. The etching was performed using an argon ion bean. The small surface conductance of the STO substrates induced by etching  (typically 1 M$\Omega$ per square) was neutralized by a one hour annealing in air at 300 $^\circ$C, which had no effect on the film properties. Magneto-conductance measurements were performed in Quantum Design \textit{ppms} system, and the zero field ones in a home-made cryostat. The geometry of the patterned samples used to measure the Hall number minimized the contribution from the longitudinal resistivity (which was found to be the one of about 0.3 \% of the Hall bar width). To further eliminate this contribution, the transverse resistivity was systematically measured by reversing the magnetic field.

\section{\label{Magneto}Magneto-Transport properties}

\subsection{Anisotropy}

\begin{figure}
\resizebox{0.95\columnwidth}{!}{%
  \includegraphics{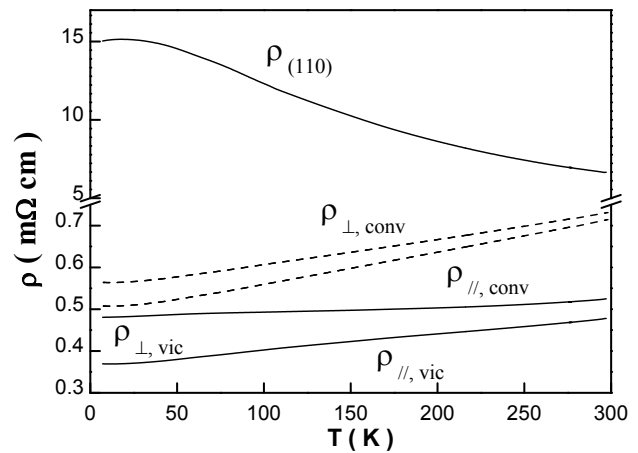}
  }
\caption{Resistivity along ($\rho_{\parallel , vic}$) and perpendicular ($\rho_{\perp, vic}$) to the film terraces, for the vicinal sample \textsl{E} (full lines) and for a conventional one (sample \textsl{C}; dashed lines). $\rho_{(110)}$ is the resistivity transverse to the film, obtained from the vicinal resistivities.}\label{vicinal}
\end{figure}

The quasi-cubic sub-structure of bulk SIO allows to anticipate essentially three-dimensional transport properties. Indeed, resistivity measurements on thin films showing in-plane orientation, as described above, only display a weak anisotropy between (001) and (1$\bar{1}$0) orientations (typically, a 2 \% difference, for a substrate with a 0.5 deg. miscut), indicating that there is no strong resistivity anisotropy between the (001) and the (110) directions. While epitaxy preserves the in-plane quasi-quadratic symmetry of the sub-structure, it likely introduces out-of-plane anisotropy, due to the monoclinic strained structure. In order to evaluate this effect, we grew films on vicinal substrates and measured their resistivity parallel and perpendicular to the dense terraces resulting from the crystal miscut. As shown in Fig.~\ref{cafm}d, substrates with a large miscut angle exhibit stacks of several elementary unit cells on each substrate terrace. The resistivity transverse to the substrate crystallographic direction is obtained using\cite{Kao93}: $\rho_{110}=(\rho_{\perp}-\rho_{\parallel} \cos^2(\theta))/ \sin^2(\theta)$, where $\rho_{\parallel}$, $\rho_{\perp}$ is the resistivity respectively parallel and perpendicular to the terraces, and $\theta$ is the vicinal angle. Doing so, we obtain that the transverse resistivity is more than one order of magnitude larger than the in-plane one, and insulating-like (Fig.~\ref{vicinal}). This could be anticipated also from the direct observation of the conductive-AFM image. Indeed, the lower resistance ribbon along the terrace step observed in Figs.~\ref{cafm}a,b (also evidenced as a resistance dip, at the terrace step in Fig.~\ref{cafm}c) is typical for the establishment of a lateral contact on the crystallographic steps, when the in-plane resistivity is lower than the transverse one\cite{Schneegans98,Degardin00}. We observe that localization effects are likely be increased in two ways, when strain modifies the band structure, as: i) the anisotropy introduces a two-dimensional character enhancing localization effects; ii) the lifting of the Dirac gap degeneracy likely goes with an increase of the carrier mass.

\begin{figure}
\resizebox{0.95\columnwidth}{!}{%
  \includegraphics{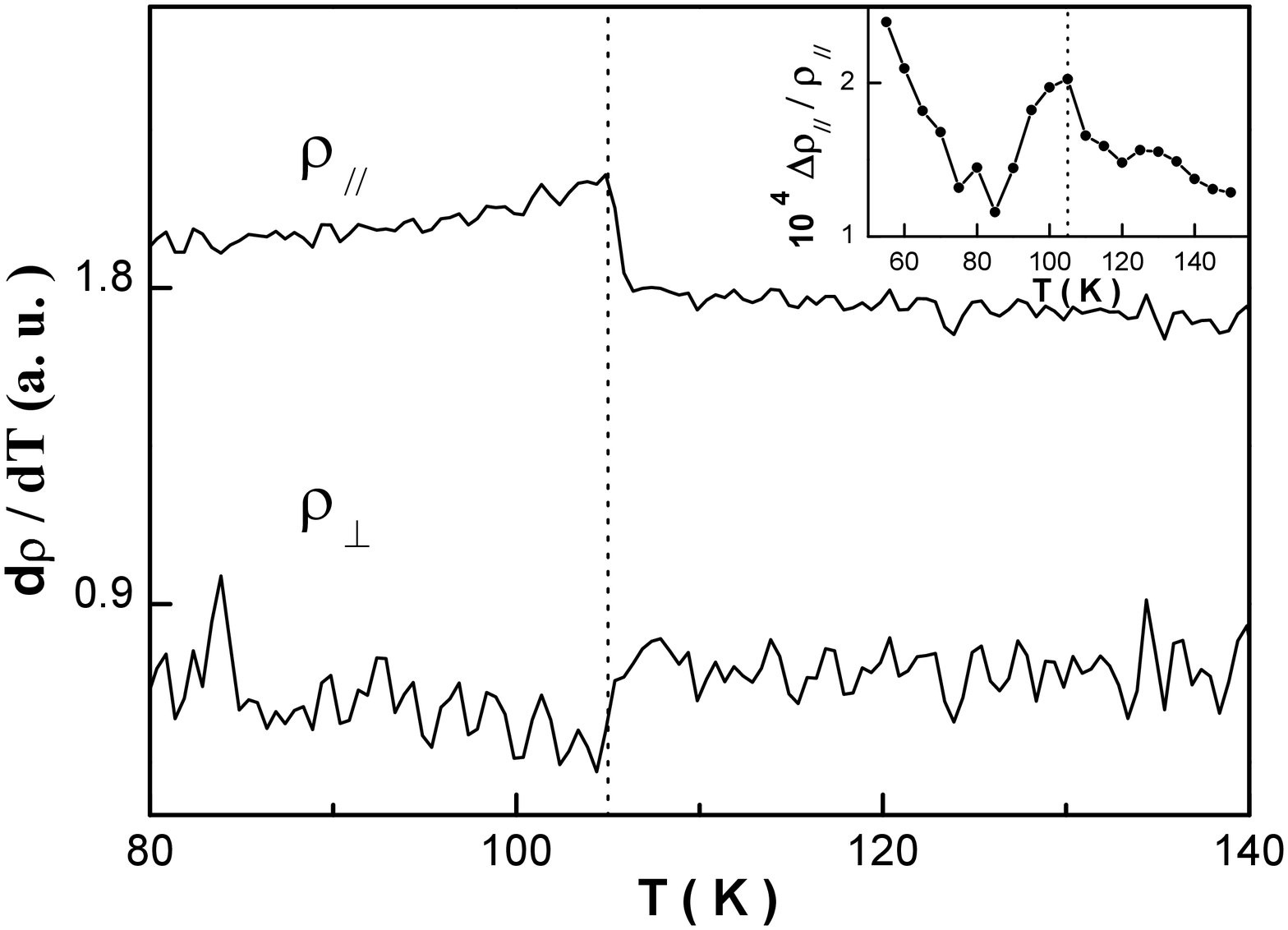}
  }
\caption{Resistivity anomaly at $T$ = 105 K, measured along ($\rho_{\parallel}$) and perpendicularly ($\rho_{\perp}$) to the miscut direction of sample \textsl{E}. The inset shows the magnetoresistance at 5 T.}\label{anomaly}
\end{figure}

A careful investigation uncovered a small, but systematic and abrupt resistivity slope jump at $T$ = $105 \pm 1 K$, for all films (we have checked, using different low temperature probes, that this is not an experimental artifact, and we did not find this anomaly for samples grown on GdScO$_3$). These low-noise measurements were performed in the home-made cryostat, using a Keithley 2402 nanovoltmeter / 2400 current source, with an unregulated temperature drift $\approx$ 0.3 K/min. The temperature for the anomaly was the same for film thickness ranging from 8 nm to 40 nm. Using a vicinal substrate, the jump was found opposite for the resistance along the terrace direction, and the one transverse to it (Fig.~\ref{anomaly}). This shows that the effect has an opposite sign for the (001) direction and the (110) one. We attribute the anomaly to the cubic to tetragonal antiferrodistortive transition in STO. This transition is characterized by the rotation of the oxygen octahedra around a cubic main axe, and by a $\approx$ $5 \, 10^{-4}$ variation of the lattice parameters\cite{Lytle64}. We expect that, due to the epitaxial strain, the $\textit{c}$-axis of the low temperature tetragonal phase lies in the plane of the substrate. Then, the increase of the STO in-plane parameter induces a small relaxation of the epitaxial strain, which results in an increase of the metallicity, as found in Fig.~\ref{anomaly}. Using a simple linear fit of the data for $T \, >$ 105 K, we estimate that there is a 15 \% resistivity change for a 1\% lattice parameter change. This is precisely what is obtained from the comparison of the resistivities of 35 nm films grown on STO and DyScO$_3$\cite{Biswas14}. Considering that the substrate transition does not alter disorder in the film, this shows that the resistivity variation observed on two such substrates at 105 K is likely not due to different amounts of disorder on different substrates, but intrinsically to the SIO unit cell deformation (which does not preclude a contribution from disorder, but the evolution of the transport properties with strain may be thought as realized at a fixed disorder strength).

\subsection{Magnetoconductance}

\begin{figure}
\resizebox{0.95\columnwidth}{!}{%
  \includegraphics{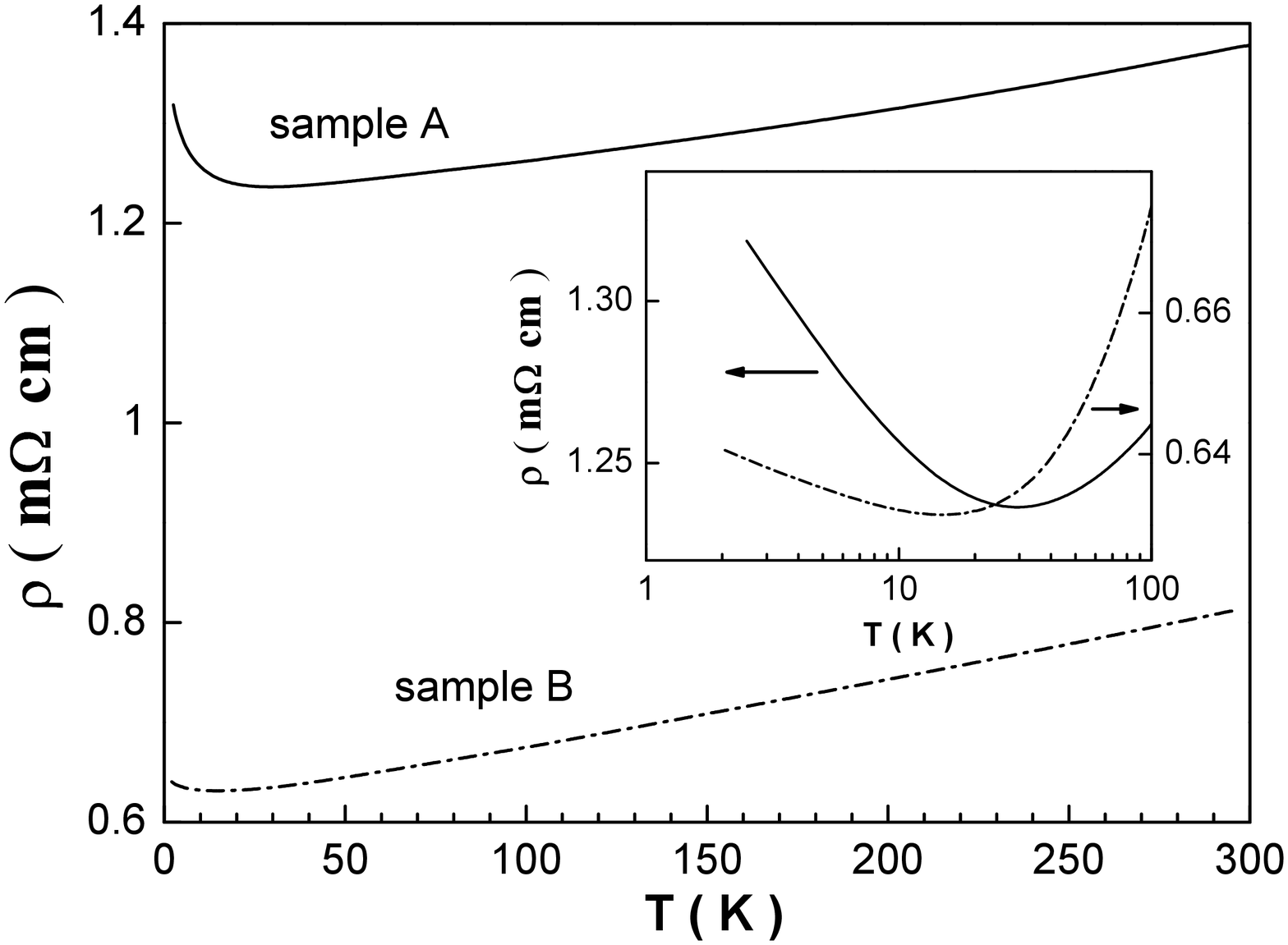}
  }
\caption{Resistivities for the 8 nm (\textsl{A}) and 14 nm (\textsl{B}) thick samples. The inset shows the $\log(T)$ behavior in the weakly localized regime.}\label{Rau}
\end{figure}

\begin{figure}
\resizebox{0.95\columnwidth}{!}{%
\includegraphics{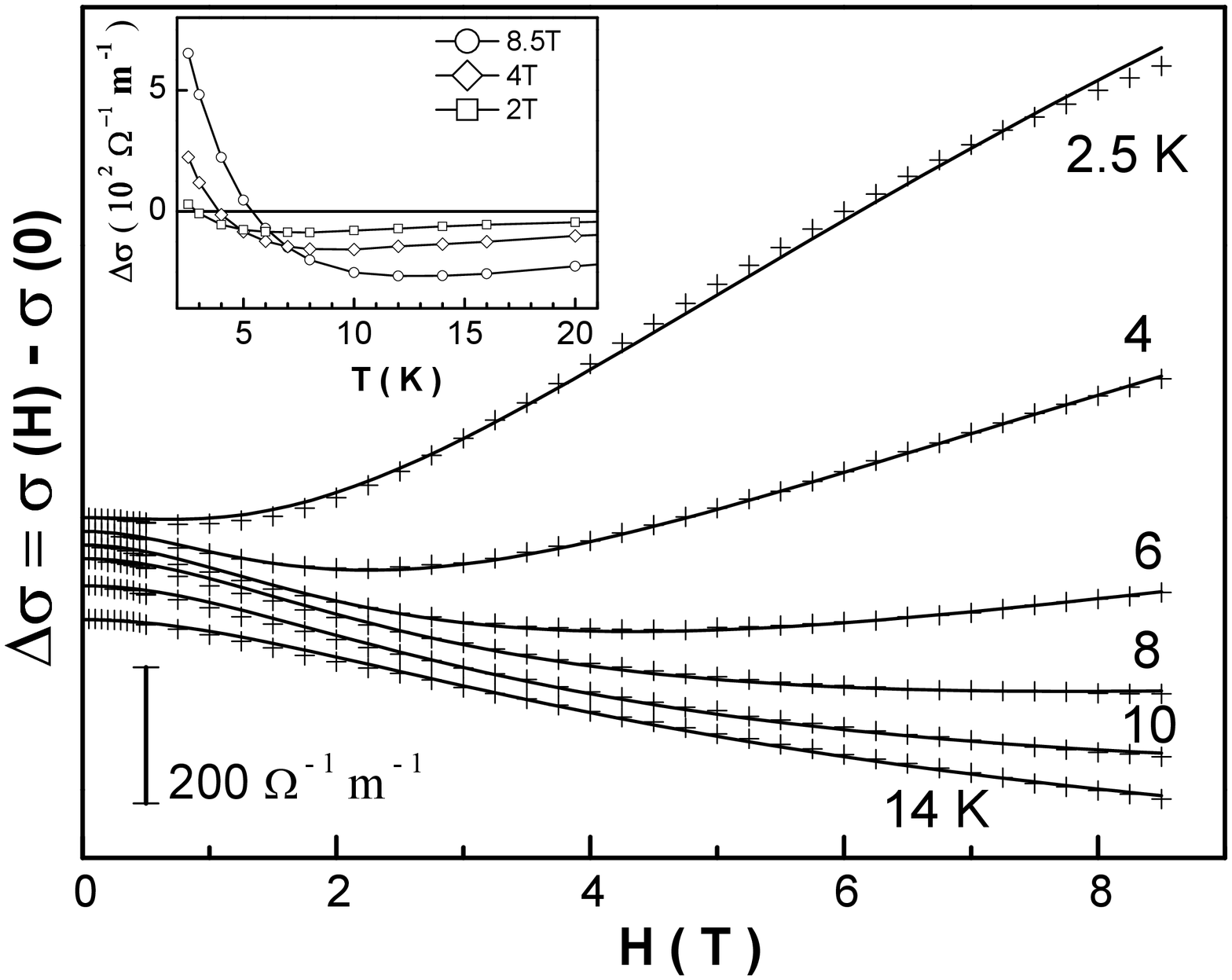}
  }
\caption{Magnetoconductance for sample \textsl{A} (field perpendicular to the film - data sets are shifted, for clarity). The lines are fits with Eqs.~\ref{MF},\ref{His}, using $H_0$ = 2.3 T; $H_{so}$ = 0.8 T; $H_i$ and $H_s$ as given in Fig.~\ref{param}.}\label{MR8nm}
\end{figure}

\begin{figure}
\resizebox{0.95\columnwidth}{!}{%
  \includegraphics{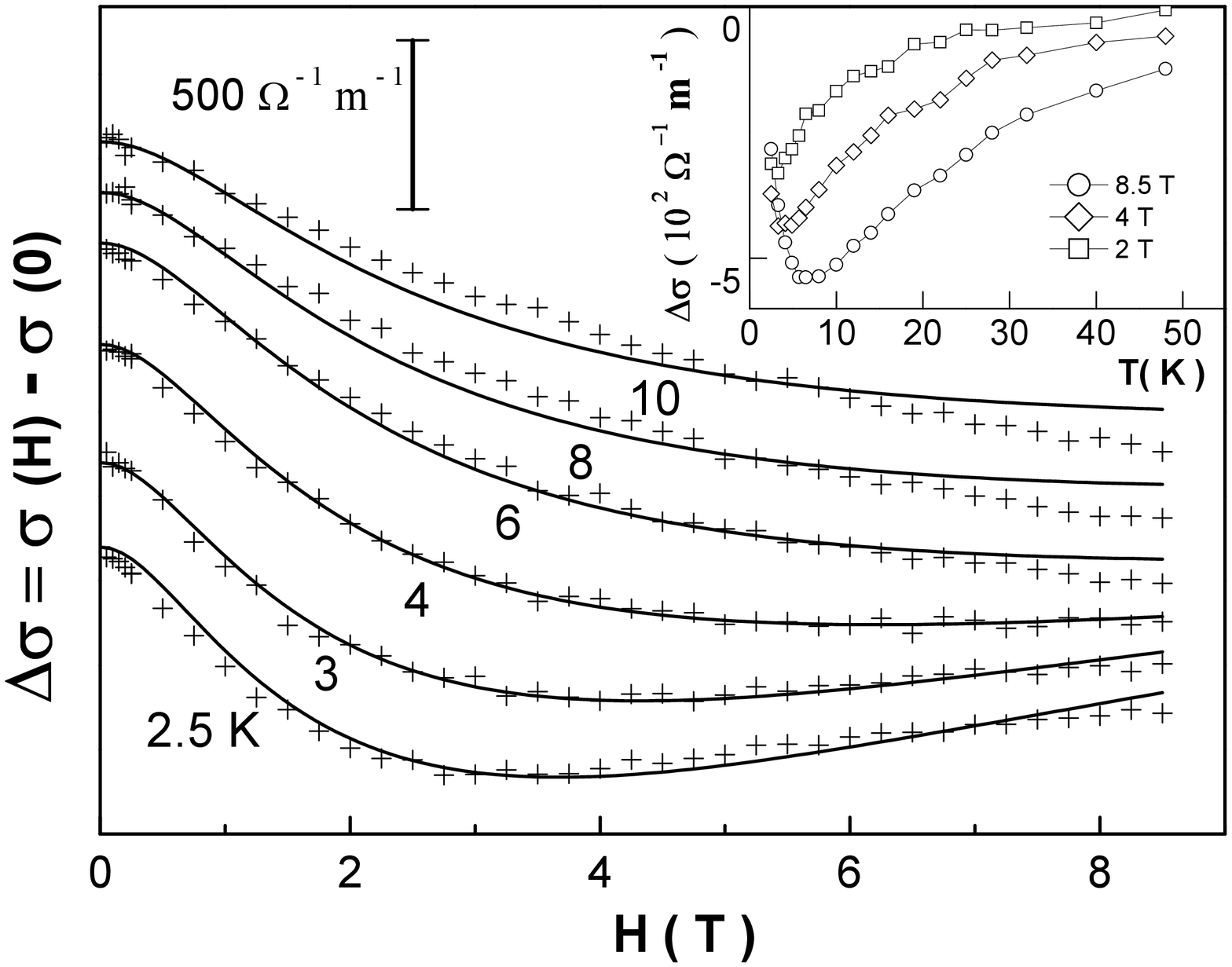}
  }
\caption{Magnetoconductance for sample \textsl{B} (field perpendicular to the film - data sets are shifted, for clarity). The lines are fits with Eqs.~\ref{MF},\ref{His}, using $H_0$ = 1.7 T; $H_{so}$ = 0.8 T; $H_i$ and $H_s$ as given in Fig.~\ref{param}.}\label{MR14nm}
\end{figure}

The resistivity was found metallic for all films, with a low temperature upturn. For the 8 nm and the 14 nm thick SIO films, the upturn shows up below $T_{min}$ =  30 K and $T_{min}$ = 15 K, respectively (Fig.~\ref{Rau}). At low temperature, $\rho(T) \sim \log(T)$, so we interpret the resistivity upturn as the manifestation of weak localization in reduced dimension. The slope for the sheet conductance from Fig.~\ref{Rau} is larger by 50\% than the theoretical value, $e^2/2\pi^2\hbar$. This could originate from the proximity of a three dimensional regime for particle diffusion. As is well known, a magnetic field destroys the constructive interference of the scattered waves of an electron at an impurity center, and thus a \textit{positive} magnetoconductance results, which depends on the relative magnitudes of the elastic scattering time and of the temperature-dependent inelastic one. This is indeed observed for the low-temperature / high-field limit for the thinner sample (Fig.~\ref{MR8nm}). In the presence of spin-orbit coupling, the scattered electron waves undergo a spin rotation which, in zero field, destroys the constructive interference and the weak localization. As the spin rotation is smaller in a strong magnetic field, this anti-localization effect is weaker in a larger field and yields a \textit{negative} magnetoconductance (for a review, see Ref.~\onlinecite{Bergmann84}). To these two antagonistic effects, one must add the contribution of magnetic impurities. As underlined in Ref.~\onlinecite{Bergmann82}, the latter is experimentally hardly distinguished from the inelastic and spin-orbit contributions.

The expression of the conductance for a two-dimensional system in a perpendicular field is, neglecting the Zeeman splitting, as well as the scattering anisotropies\cite{Hikami80, Maekawa81}:

\begin{equation}
-\sigma(H) \propto \Psi(\frac{1}{2}+\frac{H_1}{H})-\Psi(\frac{1}{2}+\frac{H_2}{H})+\frac{1}{2}\Psi(\frac{1}{2}+\frac{H_3}{H})-\frac{1}{2}\Psi(\frac{1}{2}+\frac{H_4}{H})
\label{MF}
\end{equation}

where, neglecting corrections of the order of $(H_{so}/H_0)^2$, the characteristic fields are given by\cite{Bergmann82}:

\begin{eqnarray}
H_1=H_0+H_{so}+H_s \nonumber \\
H_2=H_4=\frac{4}{3}H_{so}+\frac{2}{3}H_s+H_i \nonumber \\
H_3=2 H_s + H_i 
\label{His}
\end{eqnarray}

The characteristic fields ($H_i$, $H_0$, $H_{so}$, $H_s$; respectively: inelastic, elastic, spin-orbit coupling and magnetic scattering) are related to the scattering times ($\tau_i$) according to: $H_i$ = $\hbar/4eD\tau_i$, where $D$ is the diffusion coefficient. The critical field, $H_i$, was discussed for instance in Ref.~\onlinecite{Bergmann82}, and, whatever the dominant mechanism for the inelastic scattering, is an increasing function of the temperature. $H_0$ and $H_{so}$ should be taken in a first approximation as \textit{independent} of the temperature (see e.g. Refs.~\onlinecite{Bergmann82,Koike85} for examples on simple metals), as we did when fitting the data, using Eq.~\ref{His}. We stress that it is not, in general, possible to determine a unique set of fitting parameters, without assuming that a single set for $H_0$ and $H_{so}$ must be used to fit data at several temperatures. 

At low temperature, when $H_i \ll H_{so}$, and with no magnetic scattering ($H_s = 0$), the spin-orbit coupling contribution dominates, and the magnetoconductance is always \textit{negative}. With increasing $H_i$, the magnetoconductance obtained from Eqs.~\ref{MF} and \ref{His} either increases monotonously or displays a maximum. Thus, it is not possible to account for the data in Fig.~\ref{MR8nm} (inset), which shows opposite behavior, using $H_s$ = 0 and a monotonously increasing $H_i(T)$. Introducing a finite value for $H_s$ essentially shifts the origin of temperature, and cuts the low temperature negative divergence of the magnetoconductance. However, it was still not possible to fit the data using a temperature-independent magnetic scattering: the fits were not satisfying over the whole temperature range in Fig.~\ref{MR8nm}, and $H_i(T)$ was found an unphysical, non-monotonous, function of the temperature. Thus, we have to conclude that a temperature-dependent magnetic scattering must be introduced to fit the data, within our fitting hypothesis.

\begin{figure}
\resizebox{0.95\columnwidth}{!}{%
  \includegraphics{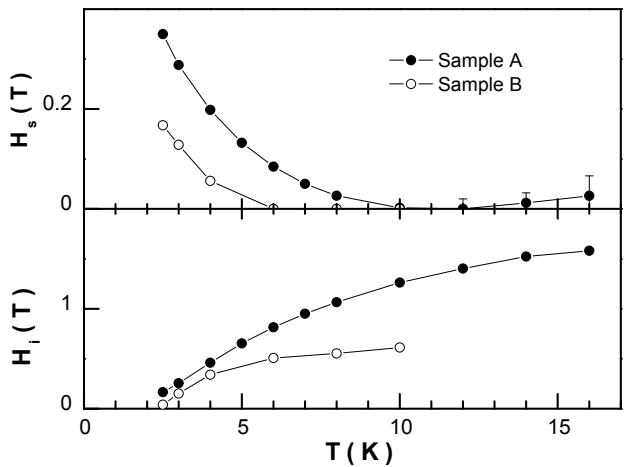}
  }
\caption{Inelastic and magnetic scattering parameters obtained from the fits in Fig.~\ref{MR8nm} (line is guide to the eye).}\label{param}
\end{figure}

Using temperature-independent $H_0$ and $H_{so}$ and allowing temperature-dependent values for $H_s$ and $H_i$, we found that best fit of the magnetoconductance curves in the temperature interval 2.5 K - 16 K is obtained using $H_{so}$= 0.7 T, $H_{0}$ = 2.3 T, and $H_s$, $H_i$ as displayed in Fig.~\ref{param}. The magnetoconductance for the thicker sample (Fig.~\ref{MR14nm}) could not be used to infer the all scattering parameters in a similar way: this is due to a larger inelastic scattering (compared to the other terms), which results in an almost monotonous, negative, magnetoconductance for the accessible field range. However, assuming $H_{so}$ is the same as for sample \textsl{A}, $H_o$ and $H_s$ are indeed obtained both smaller than for sample \textsl{A} (Fig.~\ref{param}). In a general way, due to the latitude in the fitting parameters, we believe that the temperature-dependence for $H_i(T)$ cannot be used to obtain any reliable information on the mechanism for inelastic scattering (a difficult task, even for model systems - see Ref.~\onlinecite{Bergmann84}). In particular, the saturation for $H_i(T)$, for $T \gtrsim$ 8 K, is unexpected. The reason might be that, as the inelastic lifetime of the carriers decreases when temperature rises, and becomes shorter than the diffusion time transverse to the film, the system enters the 3D regime. Still, the need for a temperature-dependent $H_s(T)$, indicating a decrease of the magnetic scattering time at lower temperature, cannot be overlooked. In particular, $H_s(T)$ seems to diverge, as $T \rightarrow 0$, which would not be expected if the need for a finite magnetic scattering was due to a correction from the actual dimensionality. A possible explanation for this behavior could be the existence of paramagnetic scattering centers, with a Curie-like effective moment. We note that it was proposed that local magnetic moments could be the source, close to the metal-insulator transition, for strong inelastic scattering centers (Ref.~\onlinecite{Biswas16}). An alternative in Ref.~\onlinecite{Zhang14} is to use a temperature-dependent SOC, originating from an unusual temperature dependence of the Land\'{e} factor. This is problematic, as the variation has to be 1-2 orders of magnitude larger than usually observed (it was suggested, however, based on comparisons with gaped semiconductors, that the Coulomb interaction variation with temperature -- see below -- may account for this unusual magnitude\cite{Zhang14}). Also, the inelastic scattering time is found, in this case, independent of the temperature, which is also rather unusual.

\subsection{Hall number}

\begin{figure}
\resizebox{0.95\columnwidth}{!}{%
  \includegraphics{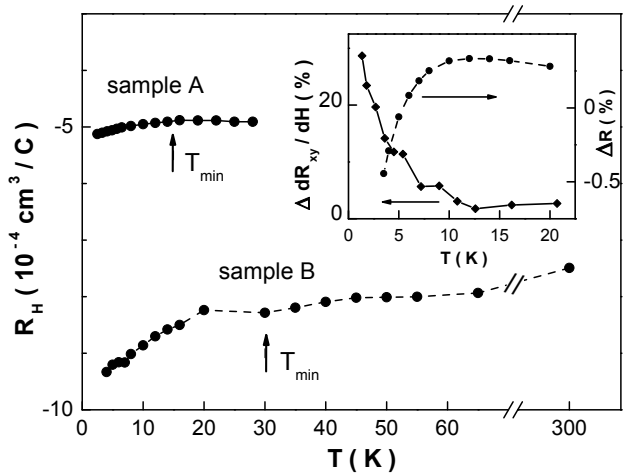}
  }
\caption{Hall coefficient for the 8 nm (\textsl{A}) and the 14 nm (\textsl{B}) thick films ($H$ = 8.5 T). The arrow indicates the temperature for the minimum in $\rho(T)$. The inset shows the non-linearity of the transverse resistance, $\Delta \,dR_{xy}/dH$ = $(dR_{xy}/dH)_{8.5\, T}-(dR_{xy}/dH)_{1.5\, T}$, and the longitudinal magnetoresistance, $\Delta \,R$, at 8.5 T.}\label{RH}
\end{figure}

The transverse resistivity was found almost linear with the applied magnetic field for the 14 nm thick sample, and non-linear at low field and low temperature for the 8 nm thick one (Fig.~\ref{RH}). Such a non-linear behavior was reported in Ref.~\cite{Zhang15} for a 7 nm thick film on STO, although stronger than observed here (Ref.~\cite{Zhang15} reports a 20\% variation between $H$ = 1 T and $H$ = 5 T, while a 14 \% one is observed in the present case. This was interpreted within an equal-mobility two-carrier model, with a weak carrier density asymmetry ($\simeq 10^{-3}$) The linear behavior for the thicker film is similar to the observations in Ref.~\onlinecite{Liu13}. However, we do not observe a strong temperature dependence for the Hall coefficient as reported in this work: as shown in Fig.~\ref{RH}, above the temperature for the minimum resistivity $\rho(T)$, when the contribution of weak localization becomes negligible, $R_H(T)$ is here fairly constant. Also, our thick film exibits a 30 times smaller $R_H$ than for the 18-20 nm thick film on STO in Ref.~\onlinecite{Liu13}, while the thinner film has a comparable Hall coefficient as reported in Ref.~\onlinecite{Zhang15}. Within the interpretation in Ref.~\onlinecite{Zhang15}, the strong temperature dependence in Ref.~\onlinecite{Liu13} and the comparatively large Hall coefficient would suggest a strong carrier asymmetry, while it could be smaller here, as well as in Ref.~\onlinecite{Zhang15}. 

However, the transverse resistivity non-linearity appears as being correlated to the weak localization magnitude, as well as the value of the Hall coefficient itself. Indeed, as seen in Fig.~\ref{RH} (inset), the temperature range where a strongly non-linear Hall signal is observed ($T \lesssim$ 10 K) coincides with the one where the anti-localization effect of the magnetic field dominates the magnetoresistance, and the Hall coefficient exhibits a marked decrease as one enters the localized regime. Thus, the principle that weak localization should not alter the Hall coefficient is here defied.

As early shown, one may get round this principle in the presence of Coulomb interaction\cite{Altshuler80,Song94} (for a review, see Refs~\onlinecite{Bergmann84,Lee85}). Indeed, the electronic interaction brings a correction for both the localization conductance change (a correction factor in 2D, in a first approximation), and for the Hall coefficient. To separate the interaction contribution (or 'Coulomb anomaly') from the conventional one, localization may be suppressed with a magnetic field (see e.g. Refs.~\onlinecite{Bishop80,Komnik83}). We may evaluate the contribution of the interaction in a similar way here also. At 4 K, data in Fig.~\ref{RH} indicates that there is a $\simeq$ 20 \% anomalous contribution ($\delta R_H$) to the Hall number. A similar estimate is also obtained from the decrease of $R_H$ from its high temperature asymptotic value in Fig.~\ref{RH}. At this temperature, neglecting the small magneto-resistance, the low-temperature resistance increase due to localization, using a high-temperature fit of the resistivity data in Fig.~\ref{Rau}, is evaluated $\delta R/R \simeq$ $7\,10^{-2}$. Thus, we have $\delta R_H/R_H \simeq$ 2.8 $\delta R/R$. This is reasonably close to the theoretical prediction, $\delta R_H/R_H \simeq 2\, \delta R/R$, considering the lack of precision for the extrapolation of the high-temperature resistivity to the localized regime. For the 14 nm thick sample, we have, at 4 K, $\delta R_H/R_H \simeq$ 4\% and $\delta R/R \simeq$ $2.5\,10^{-2}$, which also reasonably agrees with theory. However, the Hall data is not precise enough to quantify in this case the non-linearity, which is expected 5 times smaller than for the 8 nm thick sample.

This shows that both the low-temperature decrease for $R_H$ and its field-dependence may be reasonably attributed to the interaction contribution to localization. Within this interpretation, it is found that the conductance change at low temperature is essentially brought by this channel, thus underlining the importance of correlations in the present case.

\section*{Conclusion}

In conclusion, we have found that epitaxially strained SrIrO$_3$ thin films on SrTiO$_3$ exhibit several peculiar transport magneto-transport properties. The unit cell deformation, likely towards a strongly distorted monoclinic cell, is at the origin of the large anisotropy observed between in and out-of-plane directions. The resistivity anomaly associated to the substrate structural transition, at $T$= 105 K, allows to estimate the contribution of strain to resistivity, at fixed disorder. We propose, in the analysis of the magnetoconductance data within conventional weak localization theory, that the use of a temperature-dependent magnetic term, originating from localized moments on scattering centers, could be an alternative to an anomalously large temperature dependence of the Land\'{e} factor. Finally, the field and temperature dependences of the Hall data are interpreted as the consequence of a dominant anomalous contribution to the Hall effect, as a result of Coulomb interaction of carriers for this correlated semimetal.

\section*{}

We acknowledge support from the Agence Nationale de la Recherche grant SOCRATE, and valuable help from G. Collin in the handling of crystallographic data.


\end{document}